\documentclass[a4paper,11pt]{article}
\pdfoutput=1
\usepackage{jheppub}

\usepackage{amssymb,amsmath,amsfonts}
\usepackage[normalem]{ulem}
\usepackage[utf8x]{inputenc}
\usepackage{slashed}
\usepackage{graphicx}
\usepackage{tabularx}
\usepackage{here}
\usepackage{color}
\usepackage{csquotes} 
\usepackage{comment}
\usepackage{mathrsfs}
\usepackage{float}
\usepackage{ascmac}
\usepackage{multirow}
\usepackage{longtable}
\usepackage{bm}
\usepackage{ulem}

\usepackage[italicdiff]{physics}

\makeatletter
\newcommand*\rel@kern[1]{\kern#1\dimexpr\macc@kerna}
\newcommand*\widebar[1]{%
  \begingroup
  \def\mathaccent##1##2{%
    \rel@kern{0.8}%
    \overline{\rel@kern{-0.8}\macc@nucleus\rel@kern{0.2}}%
    \rel@kern{-0.2}%
  }%
  \macc@depth\@ne
  \let\math@bgroup\@empty \let\math@egroup\macc@set@skewchar
  \mathsurround\z@ \frozen@everymath{\mathgroup\macc@group\relax}%
  \macc@set@skewchar\relax
  \let\mathaccentV\macc@nested@a
  \macc@nested@a\relax111{#1}%
  \endgroup
}
\makeatother

\numberwithin{equation}{section}

\preprint{
\begin{minipage}{5cm}
\small
\flushright
EPHOU-25-017\\
KYUSHU-HET-339
\end{minipage}}

\title{
Non-Invertible Selection Rules on Heterotic Non-Abelian Orbifolds
}

\author{Tatsuo Kobayashi$^{1}$,} 
\author{Ryusei Nishida$^{1}$, and } 
\author{Hajime Otsuka$^{2}$}
\affiliation{
$^1$Department of Physics, Hokkaido University, Sapporo 060-0810, Japan\\
$^2$Department of Physics, Kyushu University, 744 Motooka, Nishi-ku, Fukuoka 819-0395, Japan
}
\emailAdd{kobayashi@particle.sci.hokudai.ac.jp}
\emailAdd{r-nishida@particle.sci.hokudai.ac.jp}
\emailAdd{otsuka.hajime@phys.kyushu-u.ac.jp}

\abstract{
We investigate coupling selection rules in heterotic string theory on non-Abelian orbifolds. 
Since boundary conditions on the orbifolds are classified by conjugacy classes of space group elements, non-Abelian orbifolds give rise to non-invertible  selection rules on couplings among twisted sectors as well as ones including untwisted sectors. Furthermore, we find that non-invertible selection rules lead to characteristic patterns of Yukawa matrices.
}

\makeatletter
\gdef\@fpheader{}
\makeatother

\begin{document}

\maketitle

\section{Introduction}
\label{sec:Intro}

String theory is a promising candidate for a unified theory of particle physics.
We can derive four-dimensional (4D) low energy effective field theory from string theory when we fix the geometry of six-dimensional compact space as well as the gauge background.
The coupling selection rules are important in 4D low energy effective field theory, and can be derived by stringy coupling rules.
Some of the stringy coupling selection rules can be understood by group theory, but we have stringy coupling rules, which can not be understood by group theory.
For example, in Refs.~\cite{Kobayashi:2024yqq,Funakoshi:2024uvy}, it was shown that orbifold compactifications with magnetic flux backgrounds lead to 
non-invertible coupling selection rules.
Such non-invertible coupling selection rules, i.e., $\mathbb{Z}_2$ gauging of $\mathbb{Z}_N$ symmetries, were applied to phenomenology of particle physics and interesting results have been obtained such as novel quark and lepton mass textures~\cite{Kobayashi:2024cvp,Kobayashi:2025znw,Kobayashi:2025ldi,Kobayashi:2025cwx,Nomura:2025sod,Chen:2025awz,Okada:2025kfm,Jangid:2025krp}, axion-less solutions of the strong CP problem~\cite{Liang:2025dkm,Kobayashi:2025thd} and other aspects~\cite{Suzuki:2025oov,Kobayashi:2025lar,Suzuki:2025bxg}.
(See, e.g. Refs.~\cite{Gomes:2023ahz,Schafer-Nameki:2023jdn,Bhardwaj:2023kri,Shao:2023gho}, for reviews on non-invertible symmetries.)\footnote{See, e.g. Refs.~\cite{Choi:2022jqy,Cordova:2022fhg,Cordova:2022ieu,Cordova:2024ypu,Delgado:2024pcv}, for other applications of non-invertible symmetries in particle physics.}

Heterotic orbifold models are one interesting type of string compactification models \cite{Dixon:1985jw,Dixon:1986jc}. 
Since one can solve analytically string theory on orbifolds, heterotic orbifold models can lead to realistic massless spectra such as gauge groups of the standard model, its extensions and three generations of quarks and leptons \cite{Ibanez:1986tp,Ibanez:1987sn,Kobayashi:2004ud,Kobayashi:2004ya,Buchmuller:2005jr,Buchmuller:2006ik,Lebedev:2006kn,Lebedev:2007hv}.
Yukawa couplings and higher order couplings were studied in Refs.~\cite{Hamidi:1986vh,Dixon:1986qv,Burwick:1990tu,Choi:2007nb}.

Stringy selection rules were also studied on Abelian orbifolds.
Some of stringy rules can be understood by group theoretical symmetries, e.g. R-symmetry \cite{Font:1988nc,Kobayashi:2004ya,CaboBizet:2013gns,Nilles:2013lda,CaboBizet:2013hms} and flavor symmetries such as $\mathbb{Z}_N$, $D_4$ and $\Delta(54)$ symmetries \cite{Dijkgraaf:1987vp,Kobayashi:2004ya,Kobayashi:2006wq,Beye:2014nxa}.\footnote{Recently, this $D_4$ symmetry was studied from the viewpoint of non-invertible symmetry \cite{Thorngren:2021yso,Heckman:2024obe,Kaidi:2024wio}.}
However, one can not understand all of the stringy coupling selection rules only by group theory.
The boundary condition of closed string on an orbifold is written by a space group element.
Conjugates of the space group element also lead to the same boundary condition of a closed string.
Thus, the boundary condition of a closed string on an orbifold is specified by a conjugacy class of the space group \cite{Hamidi:1986vh,Dixon:1986qv}.
That leads to non-invertible coupling selection rules in a certain case \cite{Kobayashi:1990mc,Kobayashi:1991rp,Kobayashi:1995py}.
Furthermore, there are other stringy coupling selection rules, which can not be understood by group theory \cite{Cvetic:1987qx,Font:1988nc,Kobayashi:2011cw}.\footnote{Recently, coupling selection rules in heterotic string theory on Calabi-Yau manifolds were studied \cite{Dong:2025pah}.}

Here, we consider heterotic string on non-Abelian orbifolds, which is defined by $T^n/G$, where $T^n$ denotes $n$-dimensional torus and $G$ denotes a non-Abelian discrete group.
Heterotic non-Abelian orbifold models are also interesting string compactifications similar to heterotic Abelian orbifold models \cite{Inoue:1987ak,Inoue:1988ki,Inoue:1990ci,Konopka:2012gy,Fischer:2012qj,Fischer:2013qza,Funakoshi:2025lxs,Hernandez-Segura:2025sfr}. 
In this paper, we study coupling selection rules on non-Abelian orbifolds by boundary conditions of closed strings, i.e., space group elements.
These coupling selection rules are controlled not by multiplication rules of group elements of $G$, but by multiplication rules of conjugacy classes of $G$.
That leads to non-invertible coupling selection rules.

This paper is organized as follows.
In section \ref{sec:Abelian}, we revisit coupling selection rules of closed strings on Abelian orbifolds by space groups.
In section \ref{sec:non-Abelian}, we study coupling selection rules on non-Abelian orbifolds by using $T^2/S_3$, $T^6/S_3$ and $T^6/T_7$ orbifolds as examples.
Section \ref{sec:con} is devoted to conclusions.

\section{Abelian orbifolds}
\label{sec:Abelian}

We start with closed strings on toroidal compactification $T^n$.
The torus $T^n$ is constructed by $R^n/\Lambda$, where $\Lambda$ is a $n$-dimensional lattice, i.e., 
$m_i e_i$ with the lattice vectors $e_i$ and integers $m_i$.
The closed string on $T^n$ satisfies the following boundary condition:
\begin{align}
X(\sigma = \pi)=X(\sigma = 0)+v,
\end{align}
with the winding vector $v=m_ie_i$.
Obviously, when we combine two strings with winding vectors $v$ and $v'$, we obtain 
the closed string with the winding vectors $v+v'$.
In other words, we denote the winding vector of the third closed string by $v''$.
Their coupling is allowed if 
\begin{align}
v+v'=v''.
\end{align}
Such a coupling selection rule corresponds to infinite cyclic groups $\mathbb{Z}^n$, which are Abelian.
The winding vector appearing in the right hand side of the above condition is unique.
We refer to this type of selection rules as {\it invertible}.
If two or more elements appear in the right hand side, we refer to it as {\it non-invertible coupling selection rules}.
One can introduce the vertex operator with the winding vector $v$ as $V_v$.
Such a state may have non-vanishing momentum, but here we omit them. 
The above selection rule can be written by the multiplication rule:
\begin{align}
V_v V_{v'} =V_{v+v'}.
\end{align}

Next, we study Abelian orbifolds $T^n/\mathbb{Z}_N$, which are division of $T^n$ by an Abelian twist $\omega$.
The twist satisfies $\omega^N=1$.
We focus on the orbifolds $T^n/\mathbb{Z}_N$, where $N$ is prime number.
On the $T^n/\mathbb{Z}_N$, the closed string satisfies the following boundary condition:
\begin{align}
X(\sigma = \pi)=gX(\sigma = 0)= \omega^k X(\sigma = 0)+v.
\label{eq:BC-abelian1}
\end{align}
We denote the space group element $g=(\omega^k, v)$, and $\omega$ corresponds to the point group element, which is $\mathbb{Z}_N$ in this case.
When we combine two strings with boundary conditions $g=(\omega^k, v)$ and $g'=(\omega^{k'}, v')$, we obtain the following 
boundary condition:
\begin{align}
X(\sigma = \pi)=gg'X(\sigma = 0)= \omega^k g'X(\sigma = 0)+v=\omega^{k+k'} X(\sigma = 0)+v+\omega^k v'.
\end{align}
That is, the production rule in the space group is written by 
\begin{align}
(\omega^k, v)(\omega^{k'}, v')=(\omega^k \omega^{k'}, v+\omega^k v')=(\omega^{k+k'}, v+\omega^k v').
\end{align}
The inverse of $(\omega^k, v)$ is $(\omega^{-k}, -\omega^{-k}v)$.
Note that the point group part is not affected by winding vectors, i.e. $\omega^k \omega^{k'}=\omega^{k+k'}$.

Another string $hX$ with $h=(\omega^{k'},v')$ can satisfy the same boundary condition as Eq.~(\ref{eq:BC-abelian1}),
\begin{align}
hX(\sigma = \pi)=ghX(\sigma = 0),
\label{eq:BC-abelain2}
\end{align}
and we can write 
\begin{align}
X(\sigma = \pi)=h^{-1}ghX(\sigma = 0).
\label{eq:BC-abelain3}
\end{align}
That implies that each closed string corresponds not to a single element $g$ of the space group, but to a conjugacy class $[g]$ of the space group, where $[g]$ is the set including $h^{-1}gh$ for all elements $h$ in the space group.
For example, for $g=(\omega^k, v)$ and $h=(\omega^{k'},v')$, we can calculate $h^{-1}gh$ as
\begin{align}
h^{-1}gh=(\omega^k,\omega^{-k'} v+(\omega^k -1)\omega^{-k'}v').
\end{align}
The conjugacy class $[g]$ includes $h^{-1}gh$ with possible $k'$ and $v'$, where $v'=m_ie_i$, i.e., the lattice $\Lambda$ itself.
For example, the conjugacy class $[(1,0)]$ includes only the identity element $(1,0)$.
For other untwisted states, i.e. $k=0$, the conjugacy classes $[(1,v)]$ include $(1,\omega^{k'}v)$ with $k'=0,1,\cdots,N-1$.
The corresponding vertex operators can be written by 
\begin{align}
U^{(0)}_v=V_v+V_{\omega v}+\cdots + V_{\omega^{N-1}v},
\end{align}
up to normalization.
These are $\mathbb{Z}_N$ invariant untwisted states.
Also, there may appear $\mathbb{Z}_N$ variant untwisted states, whose vertex operators can be written by
\begin{align}
U^{(k)}_v=V_v+e^{2\pi i k/N}V_{\omega v}+\cdots + e^{2\pi i k(N-1)/N}V_{\omega^{N-1}v},
\end{align}
up to normalization.

The multiplication rules of $U_v^{(0)}$ are non-trivial, and 
the multiplication of $U_v^{(0)}$ and $U_{v'}^{(0)}$ is written by 
\begin{align}
U_v^{(0)}U_{v'}^{(0)} = U_{v+v'}^{(0)} +U_{v+\theta v'}^{(0)} + \cdots + U_{v+\theta^{N-1} v'}^{(0)} .
\end{align}
This multiplication rule is non-invertible.
For example, for $T^n/\mathbb{Z}_2$, where $\omega v =-v$,
we obtain 
\begin{align}
\label{eq:Z2-Usector}
U_v^{(0)} U_{v'}^{(0)} = U_{v+v'}^{(0)} + U_{v-v'}^{(0)}.
\end{align}
That corresponds to $\mathbb{Z}_2$ gauging, which was studied in Refs.~\cite{Kobayashi:2024yqq,Kobayashi:2024cvp}.
This multiplication rules can be generalized for $\mathbb{Z}_N$ variant operators as
\begin{align}
U_v^{(k)}U_{v'}^{(k')} = U_{v+v'}^{(k+k')} +U_{v+\omega v'}^{(k+k')} + \cdots + U_{v+\omega^{N-1} v'}^{(k+k')} .
\end{align}

Next, we study twisted states with $(\omega^k,v)$ with $k \neq 0$.
For prime order orbifold $T^n/\mathbb{Z}_N$ with $N=$ prime, the shift vectors $(1 -\omega^{k})\Lambda$  are equivalent to $(1-\omega)\Lambda$ as shown by concrete examples later.
Thus, the conjugacy classes $[g]$ include $(\omega^k, v + (1-\omega)\Lambda)$, and they are classified by $k\,(=0,1,\cdots,N-1)$ and the vectors $v$ (mod $ (1-\omega)\Lambda)$).
Generic couplings among conjugacy classes $[g_i]=[(\omega^{k_i},v_i)]$ with $k_i \neq 0$ are allowed if they include the elements $g_i$ satisfying 
\begin{align}
\label{eq:selection-1}
\prod_i (\omega^{k_i},v_i)=(1,(1-\omega)\Lambda).
\end{align}
Let us focus on the point group selection rules.
For such purpose, we consider the couplings among twisted strings at the origin, $v_i=0$.
For these strings at the origin, each class $[(\omega^k,0)]$ includes basically a single element $(\omega^k,0)$ up to $(1-\omega)\Lambda$.
Their coupling selection rules are determined by $\mathbb{Z}_N$ symmetry.
Now, let us study the space group selection rules including twisted strings at different fixed points.
That requires 
\begin{align}
\sum_i v_i = (1-\omega)\Lambda.
\end{align}
That is finite cyclic groups because of $(1-\omega)\Lambda$, although the symmetry on toroidal compactification is an infinite one.
Explicit symmetries such as $\mathbb{Z}_{N_1}\times \mathbb{Z}_{N_2} \times \cdots$ depend on $(1-\omega)\Lambda$.
Thus, the selection rules among twisted sectors on prime order Abelian orbifolds are invertible.
We can write the multiplication rules of conjugacy classes:
\begin{align}
[(\omega^k,v)]~[(\omega^{k'},v')]=[(\omega^{k+k'},v+v'+(1-\omega)\Lambda)],
\end{align}
for $k+k'\neq 0$.
Note that the conjugacy class appearing in the right hand side is unique.
The multiplication rules can have permutation symmetries among $[(\omega^k, v)]$.
Some of them can be understood by geometrical symmetries of orbifolds \cite{Kobayashi:2004ya,Kobayashi:2006wq}.

We can study the couplings including both twisted sector and untwisted sector.
The condition for allowed couplings is the same as Eq.~(\ref{eq:selection-1}).
However, the winding vector $v$ is identified as $v+(1-\omega)\Lambda$ for the twisted sector.
That implies that if $U_v^{(0)}$ is allowed to couple, $U_{v + (1-\omega )\Lambda}^{(0)}$ are also be allowed to couple.
These coupling selection rules are non-invertible.

The twisted boundary condition (\ref{eq:BC-abelian1}) can be realized in world sheet conformal field theory by introducing twist fields $\sigma_{[(\omega^k,v)]}$ \cite{Hamidi:1986vh,Dixon:1986qv}.
The twisted string corresponds not to a single element of the space group, but to a conjugacy class.
Thus, the twist field also corresponds to the conjugacy class $\sigma_{[g]}$, and it satisfies the multiplication rules:
\begin{align}
\sigma_{[g]}\sigma_{[g']}=\sigma_{[gg']}.
\end{align}
When $[gg']=[(1,v)]$, the right hand side is replaced by the untwisted sector $[U^{(0)}_{v+(1-\omega )\Lambda}]$ with possible $(1-\omega) \Lambda$.

Let us study simple Abelian orbifolds, $S^1/\mathbb{Z}_2$ and $T^2/\mathbb{Z}_3$ for illustration.
At first, $S^1$ is the division of $R^1$ by one-dimensional lattice $\Lambda$, whose lattice vector is $e_1$.
The $\mathbb{Z}_2$ twist $\omega$ transforms $X$ to $-X$ and $e_1$ to $-e_1$.
Thus, the sublattice $(1-\omega)\Lambda$ is spanned by $2e_1$.
There are two conjugacy classes, $[(\omega,me_1)]$ with $m=0,1$.
The conjugacy classes of the untwisted sector are written by $[(1,ne_1)]$ with $n=0,1,\cdots, \infty$, and each conjugacy class includes $(1,ne_1)$ and $(1,-ne_1)$.
However, when we study the coupling selection rules including the twisted sector, 
they can be understood by introducing two sets of conjugacy classes, $[[(1,me_1)]]$ with $m=0,1$ (mod 2), i.e.,
\begin{align}
    [[(1,me_1)]]\equiv\sum_{n}[(1,(2n+m)e_1)],
\end{align}
which corresponds to 
\begin{align}
   \tilde U^{(0)}_{m} =\sum_n U^{(0)}_{(m+2n)e_1}=\sum_n \left(V_{(m+2n)e_1} + V_{(-m+2n)e_1} \right).
\end{align}
In this case, 
there are four conjugacy classes  $[(\omega^k,me_1)]$ with $k=0,1$ $m=0,1$.
Their multiplication rules are written by 
\begin{align}
\label{eq:S1/Z2}
& [(\omega,me_1)]~[(\omega,m'e_1)]=[[(1,(m+m')e_1)]] , \nonumber \\
& [[(1,me_1)]]~[(\omega,m'e_1)]=[(\omega,(m+m')e_1)],
\end{align}
where $m, m'$ and $(m+m')$ have meaning mod 2. 
The selection rule due to the point group is $\mathbb{Z}_2$.
Also the selection rules due to winding vectors have meaning mod 2, that is, another $\mathbb{Z}_2$.
Totally, there is $\mathbb{Z}_2 \times \mathbb{Z}_2$ symmetry.
In addition, there is a permutation symmetry $[(\omega,0)] \leftrightarrow [(\omega,e_1)]$. 
Then, the multiplication rules have $D_4 \simeq (\mathbb{Z}_2 \times \mathbb{Z}_2) \rtimes \mathbb{Z}_2$ symmetry.\footnote{We follow the notation of Refs.~\cite{Ishimori:2010au,Kobayashi:2022moq}. See, Ref.~\cite{Ramos-Sanchez:2018edc}, for the Abelian
symmetry resulting from the selection rule, i.e., Abelianization of the space group.}
Note that the right hand side of the first equation in Eq.~(\ref{eq:S1/Z2}) includes an infinite number of conjugacy classes in the untwisted sector.
In this sense, these multiplication rules are non-invertible.
In addition, the second equation in Eq.~(\ref{eq:S1/Z2}) can be written for 
each conjugacy class in the untwisted sector as 
\begin{align}
\label{eq:S1/Z2-2}
& [(1,(2n+m)e_1)]~[(\omega,m'e_1)]=[(\omega,(m+m')e_1)].
\end{align}
The untwisted modes with non-vanishing winding $v\neq 0$  correspond to massive modes for generic value of $S^1$ radius except a specific value such as an enhancement point.
Hence, when we discuss the couplings among massless modes, the conjugacy class $[(1,0)]$ is important and one can set $m=n=0$ in the second equation of Eq.~(\ref{eq:S1/Z2}) and Eq.~(\ref{eq:S1/Z2-2}).
However, the first equation in Eq.~(\ref{eq:S1/Z2}) shows that 
the infinite tower of the untwisted modes appears in couplings between twisted sectors.
Therefore, the infinite tower of the world sheet instanton effects contributes to the couplings of twisted sector \cite{Hamidi:1986vh,Dixon:1986qv}.

The combinations of allowed 3-point couplings among $[(1,0)]$ and $[(\omega,me_1)]$ are obtained as 
\begin{align}
  &  [(1,0)]~[(1,0)]~[(1,0)] , \nonumber \\
    &  [(1,0)]~[(\omega,me_1)]~[(\omega,-me_1)].
\end{align}
Similarly, we can obtain  higher order couplings.

For example,  suppose that three generations of left-handed and right-handed fermions correspond to $[(1,0)]$, $[(\omega,0)]$, and $[(\omega,e_1)]$, and the Higgs mode corresponds to $[(1,0)]$.
Then, we can realized the following Yukawa matrix:
\begin{align}
    Y=
    \begin{pmatrix}
        * & 0 & 0 \\
        0 & * & 0 \\
        0 & 0 & *
    \end{pmatrix},
\end{align}
where the asterisk symbol $*$ denotes allowed entries.
This is diagonal, and we can not realize flavor mixing with only this texture.
This form of Yukawa matrix is controlled by $\mathbb{Z}_2 \times \mathbb{Z}_2$ symmetry.

We can introduce the twist fields $\sigma_{[(\omega,me_1)]}$ correspond to the conjugacy class $[(\omega,me_1)]$, and they also satisfy the same multiplication rules,
\begin{align}
\sigma_{[(\omega,me_1)]}\sigma_{[(\omega,m'e_1)]}&= \tilde U^{(0)}_{(m+m')e_1}, \nonumber \\
\tilde U^{(0)}_{me_1}\sigma_{[(\omega,m'e_1)]}&=\sigma_{[(\omega,(m+m')e_1)]},
\end{align}
as well as 
\begin{align}
 U^{(0)}_{(2n+m)e_1}\sigma_{[(\omega,m'e_1)]}=\sigma_{[(\omega,(m+m')e_1)]}.
\end{align}

They have the $\mathbb{Z}_2 \times \mathbb{Z}_2$ symmetry.
Also, there is a permutation symmetry $\sigma_{[(\omega,0)]} \leftrightarrow \sigma_{[(\omega,e_1)]}$. 
Note that the coupling selection rules among only the untwisted sector are obtained as Eq.~(\ref{eq:Z2-Usector})

Similarly, we can study the $T^2/\mathbb{Z}_3$ orbifold.
Here, we use the $SU(3)$ root lattice $\Lambda_{SU(3)}$, where simple roots are denoted by $e_1$ and $e_2$ to 
construct $T^2$ as $R^2/\Lambda_{SU(3)}$.
The $\mathbb{Z}_3$ twist $\omega$ transforms 
\begin{align}\label{eq:Z3-twist}
e_1 \to e_2, \qquad e_2 \to -e_1 -e_2.
\end{align}
The $(1-\omega)\Lambda_{SU(3)}$ is spanned by $e_1-e_2$ and $e_1+2e_2$.
Note that  $2(e_1-e_2) + (e_1+2e_2)=3e_1$.
Hence, the conjugacy classes of the twisted sectors are written by 
$[(\omega^k,me_1)]$ with $k=1,2$ and $m=0,1,2$.
The conjugacy classes of the untwisted sector are  obtained by 
$[(1,v)]$, which includes $(1,v)$, $(1,\omega v)$, and $(1,\omega^2v)$.
However, they are decomposed to three sets $[[(1,me_1)]]$ with $m=0,1,2$, 
\begin{align}
    [[(1,me_1)]] =\sum_{v'} [(1,me_1 + (1-\omega)v')],
\end{align}
when we consider 
the coupling selection rules including the twisted sectors.
In this sense, there are nine conjugacy classes, $[(\omega^k,me_1)]$ with $k=0,1,2$ and $m=0,1,2$.
Their multiplication rules are written by 
\begin{align}
\label{eq:T2/Z3}
[(\omega^k,me_1)]~[(\omega^{k'},m'e_1)]&=[(\omega^{k+k'},(m+m')e_1)], \nonumber \\
[(\omega^k,me_1)]~[(\omega^{-k},m'e_1)]&=[[(1,(m+m')e_1)]], \nonumber \\
[[(1,me_1)]]~[(\omega^k,m'e_1)]&=[(\omega^k,(m+m')e_1)],
\end{align}
where $k+k' \neq 0$. 
The selection rule due to the point group is $\mathbb{Z}_3$.
Also the selection rules due to winding vectors have meaning mod 3, that is, another $\mathbb{Z}_3$.
Totally, there is $\mathbb{Z}_3 \times \mathbb{Z}_3$ symmetry.
There is a permutation symmetry of three fixed points $[(\omega^k,me_1)]$ with $m=0,1,2$, that is the permutation symmetry among 
$m=0,1,2$.
We also have an exchange symmetry between  $[(\omega,me_1)]$ and $[(\omega^{-1},me_1)]$.
Note that the right hand side of the second equation in Eq.~(\ref{eq:T2/Z3}) includes an infinite number of untwisted conjugacy classes.
The third equation in Eq.~(\ref{eq:T2/Z3}) can be written for each untwisted conjugacy class by
\begin{align}
\label{eq:T2/Z3-2}
[(1,me_1+(1-\omega)v')]~[(\omega^k,m'e_1)]=[(\omega^k,(m+m')e_1)].
\end{align}
The untwisted modes with non-vanishing shift $v\neq 0$ correspond to massive modes for generic moduli parameters except specific values.
Thus, the conjugacy class $[(1,0)]$ is important for the couplings of the untwisted massless mode.
However, the second equation in Eq.~(\ref{eq:T2/Z3}) shows that the infinite tower of untwisted modes appear in the couplings between twisted modes. 
It indicates that the infinite tower of world sheet instanton effects contributes in the couplings among twisted sectors \cite{Hamidi:1986vh,Dixon:1986qv}.

The combinations of allowed 3-point couplings among $[(1,0)]$ and $[(\omega^k,me_1)]$ are obtained as 
\begin{align}
&    [(1,0)]~[(1,0)]~[(1,0)], \nonumber \\
&    [(1,0)]~[(\omega^k,me_1)]~[(\omega^{-k},-me_1)], \nonumber \\
&    [(\omega^k,me_1)]~[(\omega^{k},m'e_1)]~[(\omega^k,(-m-m')e_1)].
\end{align}
Similarly, we can obtain higher order couplings.

For example, suppose that three generations of left-handed and right-handed fermions correspond to $[(\omega,0)]$, $[(\omega,e_1)]$, and $[(\omega,2e_1) ] $ 
and the Higgs mode corresponds to $[(1,0)] $.
Then, we can realize the following Yukawa matrix:
\begin{align}
    Y=
    \begin{pmatrix}
        * & 0 & 0 \\
        0 & 0 & * \\
        0 & * & 0
    \end{pmatrix}.
\end{align}
This Yukawa matrix is controlled by $\mathbb{Z}_3 \times \mathbb{Z}_3$.

We can introduce the twist fields $\sigma_{[(\omega^k,me_1)]}$ correspond to the conjugacy class $[(\omega^k,me_1)]$, and they also satisfy the same multiplication rules:
\begin{align}
\sigma_{[(\omega^k,me_1)]}\sigma_{[(\omega^{k'},m'e_1)]}=\sigma_{[(\omega^{k+k'},(m+m')e_1)]},
\end{align}
for $k+k' \neq 0$.
When we introduce the sum of vertex operators,
\begin{align}
    \tilde U^{(0)}_{me_1}=\sum_n U^{(0)}_{(3n+m)e_1}= \sum_n V_{(3n+m)e_1}+V_{(3n+m)\omega e_1}+V_{(3n+m)\omega^2 e_1},
\end{align}
the other multiplication rules are also written by 
\begin{align}
    \sigma_{[(\omega^k,me_1)]}\sigma_{[(\omega^{-k},m'e_1)]}&= \tilde U^{(0)}_{(m+m')e_1}, \nonumber \\
   \tilde U^{(0)}_{me_1} \sigma_{[(\omega^{k},m'e_1)]}&=\sigma_{[(\omega^k,(m+m')e_1)]},
\end{align}
as well as 
\begin{align}
    U^{(0)}_{(3n+m)e_1} \sigma_{[(\omega^{k},m'e_1)]}=\sigma_{[(\omega^k,(m+m')e_1)]}.
\end{align}

Finally, let us discuss the orbifolds $T^n/\mathbb{Z}_N$ with $N\neq$ prime. 
When $N$ is not prime,  the sublattice $(1 -\omega^{k})\Lambda$ with $k\neq 1$ is always not equivalent to $(1 -\omega)\Lambda$.
We have to modify this point when we extend our analysis to non-prime order orbifolds.
However, the selection rules due to the point group are the same even for non-prime order orbifolds.
As an illustrating model, we study the $T^2/\mathbb{Z}_4$ orbifold.
We use the $SO(4)$ root lattice $\Lambda_{SO(4) }$, where simple roots are denoted by $e_1$ and $e_2$ to construct $T^2$ as $R^2/\Lambda_{SO(4)}$.
The $\mathbb{Z}_4$ twist $\omega$ transforms 
\begin{align}
e_1 \to e_2, \qquad e_2 \to -e_1, \qquad -e_1 \to -e_2, \qquad -e_2 \to e_1.
\end{align}
The lattice $(1-\omega)\Lambda_{SO(4)}$ is spanned by $e_1-e_2$ and $2e_1$.
Thus, the $\omega$-twisted sector has two conjugacy classes, $[(\omega,me_1)]$ with $m=0,1$.
Also, the $\omega^3$-twisted sector has two conjugacy classes, $[(\omega^3,me_1)]$ with $m=0,1$.
On the other hand, the lattice $(1-\omega^2)\Lambda_{SO(4)}$ is spanned by 
$2e_1$ and $2e_2$.
Therefore, there are four conjugacy classes, $[(\omega^2,m_1e_1+m_2e_2)]$ with $m_1,m_2=0,1$.
The coupling selection rules among only $\omega$-twisted and $\omega^3$-twisted sectors correspond to Abelian selection rules as the selection rules on prime order orbifolds.
The coupling selection rules among only $\omega^2$-twisted sector also correspond to Abelian selection rules.
The situation is different for the couplings including them.
For example, we discuss the coupling among $[(\omega^k,me_1)]$, $[(\omega^k,m'e_1)]$ and $[(\omega^2,m_1e_1+m_2e_2)]$ with $k=1,3$.
This coupling is allowed when 
\begin{align}
\label{T2/Z4-1}
[(\omega^k,me_1)]~[(\omega^k,m'e_1)]~[(\omega^2,m_1e_1+m_2e_2)]=(1,(1-\omega) \Lambda_{SO(4)}),
\end{align}
where $k=1,3$.
That leads to the following condition \cite{Kobayashi:1991rp,Kobayashi:1990mc}:
\begin{align}
\label{T2/Z4-2}
m+m'+m_1+m_2={\rm even}.
\end{align}
That implies when we fix $m$ and $m'$ in the $\omega^k$-twisted sector with $k=1,3$, two conjugacy classes in the $\omega^2$-twisted sector are allowed to couple.
Thus, the coupling selection rules due to the shift part are non-trivial.
The other combinations of allowed 3-point couplings among $[(1,0)] $, $[(\omega,me_1)] $, and $[(\omega^2,m_1e_1+m_2e_2)]$ are obtained as 
\begin{align}
    & [(1,0)]~[(1,0)]~[(1,0)], \nonumber \\
        & [(1,0)]~[(\omega^k,m_1e_1)]~[(\omega^{-k},-m_1e_1)], \nonumber \\
    & [(1,0)]~[(\omega^2,m_1e_1+m_2)]~[(\omega^2,-m_1e_1+-m_2)],
\end{align}
where $k=1,3$.
These selection rules are invertible.

We can introduce the twist fields for $[(\omega^k,me_1)]$ with $k=1,3$ as $\sigma_{[(\omega^k,me_1)]}$.
Similarly, we introduce the twist fields for $[(\omega^2,m_1e_1+m_2e_2)]$ as 
$\sigma_{[(\omega^2,m_1e_1+m_2e_2)]}$.
The twist fields $\sigma_{[(\omega^2,m_1e_1+m_2e_2)]}$ with $(m_1,m_2)=(0,0)$ and $(1,1)$ are invariant under the $\mathbb{Z}_4$ twist.
However, $\sigma_{[(\omega^2,e_1)]}$ and $\sigma_{[(\omega^2,e_2)]}$ are not invariant, but they transform each other.
The $\mathbb{Z}_4$ invariant twist field is obtained by \cite{Kobayashi:1990mc}
\begin{align}
\sigma_{[(\omega^2,e_1)]+}=\sigma_{[(\omega^2,e_1)]}+\sigma_{[(\omega^2,e_2)]}.
\end{align}
Similarly, we can write $\mathbb{Z}_4$ variant twist field as 
\begin{align}
\sigma_{[(\omega^2,e_1)]-}=\sigma_{[(\omega^2,e_1)]}-\sigma_{[(\omega^2,e_2)]}.
\end{align}
In this basis, multiplication rules of $\sigma_{[\omega,me_1]}$ are written by
\begin{align}
\sigma_{[(\omega,0)]}~\sigma_{[(\omega,0)]}&=\sigma_{[(\omega,e_1)]}~\sigma_{[(\omega,e_1)]}=\sigma_{(\omega^2,0)}+\sigma_{(\omega^2,e_1+e_2)}, \nonumber \\
\sigma_{[(\omega,0)]}~\sigma_{[(\omega,e_1)]}&=\sigma_{[(\omega^2,e_1)]+}.
\end{align}
These selection rules are still non-invertible.

We summarize the coupling selection rules on Abelian orbifolds.
For prime order orbifolds, the coupling selection rules among only twisted sectors correspond to Abelian selection rules.
For non-prime order orbifolds, the point group selection rules are also Abelian, but 
selection rules due to shifts in the space group are non-invertible.
The difference among the sublattices  $(1-\omega^k)\Lambda$ leads to 
such a property on the selection rules due to the shift.
Also, the couplings including the untwisted and twisted sectors on both prime and non-prime order orbifolds are non-invertible 
in the shift part of the space group.
The sublattice $(1-\omega^k)\Lambda$ in the shift vector $v$ is important.

\section{Non-Abelian orbifolds}
\label{sec:non-Abelian}

Here, we study closed strings on non-Abelian orbifolds $T^n/G$, where $G$ is a non-Abelian discrete group. 
The boundary conditions of the closed strings are written by
\begin{align}
X(\sigma = \pi)=gX(\sigma = 0)= u X(\sigma = 0)+v,
\label{eq:BC-non-abelian1}
\end{align}
where $u$ is an element of $G$.
We denote the space group element $g=(u,v)$.
Similar to the Abelian orbifolds, strings correspond not to a space group element $g$, but to a 
conjugacy class $[g]$.
In order to study the point selection rules, we focus on the couplings among twisted strings at the origin $v=0$.
For this case, the conjugacy class $[g]$ can be written by $[(u,0)]=([u],0)$.
Thus, the conjugacy classes of $G$ are important.\footnote{
See e.g. Ref.~\cite{Dong:2025jra} for concrete examples of multiplication of conjugacy classes.}
Suppose that the multiplication rules of conjugacy classes in $G$ can be written by
\begin{align}
[u_i]~[u_j]=c_{ijk}[u_k],
\end{align}
where the conjugacy classes appearing in the right hand side are not always unique.
Then, the multiplication rules of $[g]$ are written by 
\begin{align}
[(u_i,0)]~[(u_j,0)]=([u_i],0)~([u_j],0)=(c_{ijk}[u_k],0)=c_{ijk}[(u_k,0)],
\end{align}
where the conjugacy classes appearing in right hand side are not always unique.
Thus, we can realize non-invertible selection rules on non-Abelain orbifolds.

Next, let us study the conjugacy of the space group elements.
Note that the inverse of $(u',v')$ is $(u'^{-1}, - u'^{-1}v')$.
Then, the conjugate of $(u,v)$ is obtained 
\begin{align}
(u'^{-1}, - u'^{-1}v')(u,v)(u',v')=(u'^{-1}uu',u'^{-1}v+u'^{-1}(u-1)v').
\end{align}
Then, the conjugacy class $[(u,v)]$ includes 
$(u'^{-1}uu',u'^{-1}v+u'^{-1}(u-1)v')$ with all the possible $u'$ and $v'$.
The point group part $u'^{-1}uu'$ is the conjugacy class of $G$.
In general, the conjugacy class $[(1,v)] $ in the untwisted sector includes $(1,u'v)$, i.e.
\begin{align}
\label{eq:conjugacy-class-G}
    [(1,v)]= \{ (1,u' v)  \},
\end{align}
for $u' \in G$.
The untwisted modes with non-vanishing winding $v \neq 0$ correspond to massive modes for generic moduli parameters except for specific values.
In what follows, we focus mainly on the untwisted massless mode corresponding to $[(1,0)] $.
Similarly, if the $u$-twisted sector has an unrotated subspace, its conjugacy classes $[(u,v+v')] $ are classified by an infinite tower of winding $v$ along unrotated directions, where $v'$ denotes a vector along the twisted subspace.
These $u$-twisted modes with non-vanishing $v\neq 0$ are massive for generic moduli parameters.
In what follows, we mainly focus on the $u$-twisted massless modes corresponding to $[(u,v')] $ with $v=0$.

In what follows, we study concrete examples of multiplication rules on non-Abelian orbifolds.

\subsection{$T^2/S_3$ orbifold}

Here, we study the $T^2/S_3$ orbifold.
We construct $T^2$ as $R^2/\Lambda_{SU(3)}$, which is the same as $T^2$ for $T^2/\mathbb{Z}_3$.
Then, we define the $\mathbb{Z}_3$ twist $\omega$ in the same way as Eq.~(\ref{eq:Z3-twist}).
In addition, we introduce the $\mathbb{Z}_2$ twist $\theta$ such that it transforms the basis vectors as 
\begin{align}
e_2 ~\leftrightarrow~ -e_1-e_2,
\end{align}
 while $e_1$ is invariant.
The twists $\theta$ and $\omega$ satisfy the following algebraic relations:
\begin{align}
\theta^2=1, \qquad \omega^3=1, \qquad \theta \omega \theta = \omega^2,
\end{align}
that is, $S_3$.
Note that the sublattice $(1-\omega)\Lambda_{SU(3)}$ is the same as 
$T^2/\mathbb{Z}_3$, and it is involved in the sublattice 
$(1-\theta)\Lambda_{SU(3)}$.

First, we study the conjugacy classes of point group with the vanishing shift $v=0$.
They are written by
\begin{align}
[(1,0)]&=\{ (1,0) \}, \nonumber \\
[(\omega,0)]&=\{ (\omega,0), (\omega^2,0) \} \nonumber \\
[(\theta,0)]&=\{ (\theta,0), (\theta \omega,0), (\theta \omega^2,0) \}.
\end{align}
They are the conjugacy classes of $S_3$.
Their multiplication rules are obtained as follows:
\begin{align}
[(1,0)][(u,0)] & =[(u,0)] \nonumber \\
[(\omega,0)][(\omega,0)] & =[(1,0)] + [(\omega,0)] \nonumber \\
[(\theta,0)][(\theta, 0)] &=[(1,0)]+[(\omega,0)] \nonumber \\
[(\omega,0)][(\theta,0)] &=[(\theta,0)].
\end{align}
They are commutable.
These multiplication rules are non-invertible.
There is the $\mathbb{Z}_2$ symmetry, where $[(\theta,0)]$ is $\mathbb{Z}_2$ odd, and the others are even.

Next, we study the conjugacy classes of $(u,v)$ with non-vanishing $v$.
The conjugacy classes in the untwisted sector are written by Eq.~(\ref{eq:conjugacy-class-G}).
Similar to $T^2/\mathbb{Z}_3$, it would be convenient to introduce the set of conjugacy classes: 
\begin{align}
    [[(1,me_1)]]=\sum_{u',v'} [(1,(me_1 +(1-u')v')],
\end{align}
when we discuss the couplings with the $\omega$-twisted sector.

Now, let us discuss the $\omega^k$-twisted sectors.
We find that 
\begin{align}
    (\theta,0)(\omega,m_ie_i)(\theta^{-1},0)=(\omega^2,m'_ie_i),
\end{align}
where 
\begin{align}
    m'_1=m_1-m_2,\qquad m'_2=-m_2.
\end{align}
That means that any space group element $(\omega^2,m_ie_i)$ in the $\omega^2$ twisted sector has its conjugate $(\omega,m_i'e_i)$ in the $\omega$-twisted sector.
Also note that the conjugate transformation $h^{-1}gh$ of $g=(\omega,v)$ with $h=(\theta,v')$ transforms the $\omega$-twisted sector to $\omega^2$-twisted sector.
Thus, it is enough to classify the conjugacy classes of only the $\omega$-twisted sector by conjugate transformation with $h=(\omega^k,v')$.
Similar to the $T^2/\mathbb{Z}_3$, we find that $(1-\omega)\Lambda_{SU(3)}$ is spanned by $e_1-e_2$ and $3e_1$. 
Thus, we find that the conjugacy classes including $(\omega^k,v)$ are decomposed as 
\begin{align}
[(\omega,me_1)]=\{(\omega^k, me_1 + (1-\omega)\Lambda_{SU(3)}) \},
\end{align}
where $k=1,2$ and $m=0,1,2$.

Similarly, any space group element $(\theta \omega^k,m_ie_i)$ in the $\theta \omega^k$-twisted sector with $k=1,2$ has its conjugate $(\theta,m'_ie_i)$ in the $\theta$-twisted sector.
Thus, it is enough to classify the conjugacy classes of $(\theta,m'_ie_i)$ in the $\theta$-twisted sector only by the conjugate transformation with $h=(\theta,v)$.
Note that $(1-\theta)\Lambda_{SU(3)}=n(e_1+2e_2)$, while we can write
\begin{align}
    \Lambda_{SU(3)} &= pe_1+(1-\omega)\Lambda_{SU(3)}=pe_1 + n(e_1+2e_2)+n'(e_1-e_2) \nonumber \\
& =pe_1 + n'(e_1-e_2) + (1-\theta)\Lambda_{SU(3)},
\end{align}
where $p=0,1,2$ and $n,n' \in \mathbb{Z}$.
That implies that one direction is not twisted.
The conjugacy classes are written by 
\begin{align}
    [(\theta,pe_1 +n'(e_1-e_2))].
\end{align}
Again, it is convenient to introduce the set of conjugacy classes,
\begin{align}
     [[(\theta,pe_1 )]] = \sum_{n'}  [(\theta,pe_1 +n'(e_1-e_2))],
\end{align}
when we discuss the couplings with the $\omega$-twisted sector.

Multiplication rules of conjugacy classes are obtained as 
\begin{align}
[[(1,me_1)]]~[(\omega,m'e_1]&=[(\omega, m'e_1)] \nonumber \\
[[(1,me_1)]]~[[(\theta,m'e_1]]&=[[(\theta, (m+m')e_1)]] \nonumber \\
[(\omega,me_1)]~[(\omega,m'e_1)]&=[[(1,(m+m')e_1)]] + [(\omega,(m+m')e_1)] \nonumber \\
[[(\theta,me_1)]]~[[(\theta, m'e_1)]]&=[[(1,(m+m')e_1)]]+[(\omega,(m+m')e_1)] \nonumber \\
[(\omega,me_1)]~[[(\theta,m'e_1)]]&=[[(\theta,(m+m')e_1)]].
\end{align}
We find that the point group part is the multiplication rules of $S_3$ conjugacy classes.
On the other hand, the multiplication rules of the shift part are $\mathbb{Z}_3$, when we write them by use of sets $[[(1,(m+m')e_1)]]$ and $[[(\theta,(m+m')e_1)]]$.
However, these sets include an infinite number of conjugacy classes.
Hence, these are non-invertible.
Similar to those in the previous section, we can write multiplication rules of 
each conjugacy class in the untwisted sector and $\theta$-twisted sector.
As mentioned for above, the untwisted modes with $[(1,me_1)]$ ($m\neq 0$) and the $\theta$-twisted modes with $[(\theta, pe_1+ n'(e_1-e_2))]$ ($n'\neq 0$) correspond to massive modes for generic moduli parameters.
For massless modes, $[(1,0)]$  and $[(\theta, pe_1)]$ are important, although massive modes can contribute e.g. as world sheet instanton effects.

The combinations of allowed 3-point couplings among $[(1,0)] $, $[(\omega, me_1)] $ and $[(\theta,  pe_1)]$ are obtained as 
\begin{align}
    & [(1,0) ]~[(1,0)]~[(1,0)], \nonumber \\
        & [(1,0) ]~[(\omega,me_1)]~[(\omega,-me_1)], \nonumber \\
    & [(1,0) ]~[(\theta,pe_1)]~[(\theta,-pe_1)], \nonumber \\
    & [(\omega,me_1) ]~[(\omega,m'e_1)]~[(\omega,(-m-m')e_1)].
\end{align}
Similarly, we can obtain higher order couplings.

\subsection{$T^6/S_3$ orbifold}

In this section, we focus on the $T^6/S_3$ orbifold.
We use the $\Lambda = \Lambda_2 \times \Lambda_{SU(3)} \times \Lambda_{SU(3)}$ to construct $T^6=
R^6/\Lambda $, where $\Lambda_2$ is any two-dimensional lattice.
We denote the basis vectors of $\Lambda_2$ by $e_1$ and $e_2$.
In addition, we denote simple roots of the first (second) $\Lambda_{SU(3)}$ lattice by 
$e_3$ and $e_4$ ($e_5$ and $e_6$).
The $\mathbb{Z}_3$ twist $\omega $ transforms the basis vectors as 
\begin{align}
e_3 \to -e_3-e_4 \to e_4 \to e_3, \nonumber \\
e_5 \to e_6 \to -e_5-e_6 \to e_5, 
\end{align}
while $e_1$ and $e_2$ are invariant.
The $\mathbb{Z}_2$ twist $\theta$ transforms the basis vectors as
\begin{align}
e_1 \to -e_1,\quad e_2 \to -e_2, \quad 
e_3 \leftrightarrow e_5,\quad e_4 \leftrightarrow e_6.
\end{align}
$\theta$ and $\omega$ satisfy the $S_3$ algebra.

The conjugacy classes of point group with the vanishing shift $v=0$ are the same as those on the $T^2/S_3$ orbifold, i.e.,
\begin{align}
[(1,0)]&=\{ (1,0) \}, \nonumber \\
[(\omega,0)]&=\{ (\omega,0), (\omega^2,0) \} \nonumber \\
[(\theta,0)]&=\{ (\theta,0), (\theta \omega,0), (\theta \omega^2,0) \}.
\end{align}
They are the conjugacy classes of $S_3$.
Their multiplication rules are obtained as follows:
\begin{align}
[(1,0)][(u,0)] & =[(u,0)] \nonumber \\
[(\omega,0)][(\omega,0)] & =[(1,0)] + [(\omega,0)] \nonumber \\
[(\theta,0)][(\theta, 0)] &=[(1,0)]+[(\omega,0)] \nonumber \\
[(\omega,0)][(\theta,0)] &=[(\theta,0)].
\end{align}
They are commutable and non-invertible. 
There is the $\mathbb{Z}_2$ symmetry, where $[(\theta,0)]$ is $\mathbb{Z}_2$ odd, and the others are even.

Now, let us study the conjugacy classes of space group elements with non-vanishing $v\neq 0$.
The conjugacy classes of the untwisted sector are written by Eq.~(\ref{eq:conjugacy-class-G}).
It is convenient to introduce the following set of conjugacy classes:
\begin{align}
    &[[(1,m_1e_1+m_2e_2+m_3e_3+m_5e_5)]] \nonumber \\
    &~~~~~~~~~~~~= \sum_{u',v'} [(1,(m_1e_1 + m_2 e_2+m_3 e_3 + m_5 e_5+ (1-u')v')   ],
\end{align}
where $m_1, m_2=0,1$ and $m_3,m_5=0,1,2$ when we discuss the couplings with the $\theta$-twisted sectors and $\omega$-twisted sector.

Any space group element $(\omega^2,m_ie_i)$ in the $\omega^2$-twisted sector is conjugate to a space group element $(\omega,m_ie_i)$ in the $\omega$-twisted sector by the conjugate transformation $h^{-1}gh$ with $h=(\theta,v)$.
Thus, it is enough to classify the conjugacy classes of the $\omega$-twisted sector by the conjugate transformation with $h=(\omega,v)$.
Since the sublattice structure in the $e_3-e_4$ plane and $e_5-e_6$ plane is the same as one in $T^2/\mathbb{Z}_3$, the conjugacy classes of the $\omega$ sector as well as the $\omega^2$-twisted sector are written by
\begin{align}
    [(\omega,m_1e_1+m_2e_2+m_3e_3+m_5e_5)]=&\{ (\omega,  m_1e_1+ m_2e_2+m_3e_3+m_5e_5 + (1-\omega) \Lambda_{SU(3)}^2 ),\notag\\
    &(\omega^2,  -m_1e_1- m_2e_2+m_3e_3+m_5e_5 + (1-\omega) \Lambda_{SU(3)}^2 )\} ,
\end{align}
where $m_1,m_2 \in \mathbb{Z}$ and $m_3,m_5=0,1,2$.
That is, there are nine fixed points on the four-dimensional space spanned by 
$e_i$ with $i=3,4,5,6$, while $e_1-e_2$ space is the untwisted torus.
It is convenient to introduce the set of conjugacy classes,
\begin{align}
 &    [ [(\omega,m_1e_1+m_2e_2+m_3e_3+m_5e_5)]] \nonumber \\
 &~~~~~~~~=\sum_{n_1,n_2}
       [(\omega,(2n_1+m_1)e_1+(2n_2+m_2)e_2+m_3e_3+m_5e_5)],
\end{align}
where $m_1,m_2=0,1$ and $m_3,m_5=0,1,2$.

Similarly, we can study the conjugacy classes of $\theta \omega^k$-twisted sectors.
For that purpose, it is enough to classify the conjugacy classes of $\theta$-twisted sector by the conjugate transformation $h=(\theta,v)$.
Such conjugacy classes are written by
\begin{align}
&    [(\theta,m_1e_1+m_2e_2+m_3e_3+m_4e_4)] \nonumber \\ &~~~~~~~~~~= 
    \{ (\theta, m_1e_1+m_2e_2+m_3e_3+m_4e_4+(1-\theta)\Lambda),\nonumber\\
    &~~~~~~~~~~~~~~~(\theta\omega, m_1e_1+m_2e_2+m_4e_3+m_3e_6+\omega(1-\theta)\Lambda),\nonumber\\
    &~~~~~~~~~~~~~~~(\theta\omega^2, m_1e_1+m_2e_2+m_3e_4+m_4e_5+\omega^2(1-\theta)\Lambda)  \} ,
\end{align}
where $m_1,m_2=0,1$ and the others are integers.
It is convenient to introduce the following set of conjugacy classes:
\begin{align}
      [[(\theta,m_1e_1+m_2e_2+m_3e_3)]] = \sum_{m_4}   [(\theta,m_1e_1+m_2e_2+m_3e_3+ m_4(e_3-e_4))],
\end{align}
when we study the couplings with the $\omega$-twisted sector.

Hereafter, we use the following notation:
\begin{align}
    [[(u, m_1,m_2,m_3,m_5) ]]= [[(u, m_1e_1+m_2e_2+m_3e_3+m_5e_5) ]].
\end{align}
When $u=\theta$, we have $m_5=0$.
By use of them, the multiplication rules of the conjugacy classes are as follows,
\begin{align}
    &[[(u_i,m_1,m_2,m_3,m_5)]]~[[(u_j,m'_1,m'_2,m'_3,m'_5]] \nonumber \\
& ~~~~~~~~~~=\sum_{u_k}c_{ijk}[[(u_k,m_1+m'_1,m_2+m'_2,m_3+m'_3,m_5+m_5')   ]],
\end{align}
where $u_{i,j,k}=1,\omega,\theta$, and $u_k$ must satisfy $u_i u_j=c_{ijk} u_k$. 
These multiplication rules are direct products of multiplication rules of $S_3$ conjugacy classes and $\mathbb{Z}_2\times \mathbb{Z}_2 \times \mathbb{Z}_3$.
However, note that the set of conjugacy classes $ [[(u, m_1,m_2,m_3,m_5) ]]$ include the infinite number of conjugacy classes.
In this sense, these multiplication rules are non-invertible in the winding part.
Also the point group selection rules are non-invertible because of the multiplication rules of $S_3$ conjugacy classes.
Winding modes on untwisted tori correspond to massive modes.
We can derive the coupling selection rules of massless modes by choosing $m_i=0$ in a proper way.

The combinations of allowed 3-point couplings among $[(1,0)] $, $[(\omega,m_3e_3+m_5e_5)] $, and $[(\omega, m_1e_1+m_2e_2)] $ are obtained as 
\begin{align}
    & [(1,0)]~[(1,0)]~[(1,0)], \nonumber \\
    & [(1,0)]~[(\omega,m_3e_3+m_5e_5)]~[(\omega,-m_3e_3-m_5e_5)], \nonumber \\
    & [(1,0)]~[(\theta,m_1e_1+m_2e_2)]~[(\omega,-m_1e_1-m_2e_2)], \nonumber \\
    & [(\omega,m_3e_3+m_5e_5)]~[(\omega,m'_3e_3+m'_5e_5)]~[(\omega,(-m_3-m'_3)e_3-(m_5+m'_5)e_5)].
\end{align}
Similarly, we can obtain higher order couplings.

\subsection{$T^6/T_7$ orbifold}

Here, we study the $T^6/T_7$ orbifold.
We use the $\Lambda_{SU(7)}$ lattice to construct $T^6=R^6/\Lambda_{SU(7)}$.
We denote the simple roots of $SU(7)$ by $e_i$ with $i=1,\cdots,6$.
The $\mathbb{Z}_7$ twist $\omega$ transforms the basis vectors as 
\begin{align}
e_1 \to e_2 \to e_3 \to e_4 \to e_5 \to e_6 \to -\sum_i e_i \to e_1,
\end{align}
i.e., $\omega^7=1$.
On the $T^6/\mathbb{Z}_7$ orbifold, 
the group elements $(\omega^k,v)$ are decomposed to the following conjugacy classes:
\begin{align}
[(\omega^k,me_1)],
\end{align}
where $k=1,2,\cdots,6$ and $m=0,1,\cdots,6$.

Furthermore, we introduce the $\mathbb{Z}_3$ twist $\theta$, which satisfies the following algebraic relations:
\begin{align}
\theta^{-1} \omega \theta = \omega^2,
\end{align}
where $\omega^7=1$ and $\theta^3=1$.
That is $T_7$.
Then, the space group elements $(\omega,0)$ with the vanishing shift $v=0$ are 
decomposed to 
\begin{align}
[(\omega,0)]&=\{ (\omega,0), (\omega^2, 0), (\omega^4,0)   \}, \nonumber \\
[(\omega^3,0)]&= \{  (\omega^3,0), (\omega^5, 0), (\omega^6,0)   \}.
\end{align}
In addition, the space group elements $(\theta^k \omega^\ell,0)$ are 
decomposed to
\begin{align}
[(\theta,0)]&=\{(\theta\omega^k, 0)   \}  \nonumber \\
[(\theta^2,0)]&=\{(\theta^2\omega^k, 0)   \} ,
\end{align}
where $k=0,1,\cdots, 6$.
Their multiplication rules are ones of $T_7$ conjugacy classes, i.e.,
\begin{align}
[(1,0)]~[(u,0)]&=[(u,0)], \nonumber \\
[(\omega,0)]~[(\omega,0)]&=[(\omega,0)]+[(\omega^3,0)],  \nonumber \\
[(\omega,0)]~[(\omega^3,0)]&=[(1,0)]+[(\omega,0)]+[(\omega^3,0)],  \nonumber \\
[(\omega^3,0)]~[(\omega^3,0)]&=[(\omega,0)]+[(\omega^3,0)], \nonumber \\
[(\omega^k,0)]~[(\theta^\ell,0)]&=[(\theta^\ell,0)],  \nonumber \\
[(\theta,0)]~[(\theta,0)]&=[(\theta^2,0)],  \nonumber \\
[(\theta^2,0)]~[(\theta^2,0)]&=[(\theta,0)],  \nonumber \\
[(\theta,0)]~[(\theta^2,0)]&=[(1,0)]+[(\omega,0)]+[(\omega^3,0)]  .
\end{align}

The conjugacy classes with non-vanishing shifts $v \neq 0$ in the $\omega$-twisted and $\omega^3$-twisted sectors are written by
\begin{align}
 [(\omega,me_1)] &= \{ (\omega^k,me_1)  \}  ~~~~(k=1,2,4),  \nonumber \\
 [(\omega^3,me_1)] &= \{ (\omega^k,me_1)  \}  ~~~~(k=3,5,6) .
\end{align}
On the other hand, the $\theta$-twisted and $\theta^2$-twisted sectors have
the untwisted torus.
Up to this untwisted direction, each of the $\theta$-twisted and $\theta^2$-twisted sectors has a single fixed point \cite{Fischer:2012qj,Fischer:2013qza}.
Winding modes along the untwisted direction correspond to massive modes for generic moduli parameters.
Thus, we consider the couplings including only $[(1,0)]$, $[(\theta,0)]$ $[(\theta^2,0)]$ as well as $[(\omega,me_1)]$ and $[(\omega^3,me_1)]$.
Their multiplication rules are written by 
\begin{align}
[(1,0)]~[(u,me_1)]&=[(u,me_1)], \nonumber \\
[(\omega,me_1)]~[(\omega,m'e_1)]&=[(\omega,(m+m')e_1)]+[(\omega^3,(m+m')e_1)],  \nonumber \\
[(\omega,me_1)]~[(\omega^3,m'e_1)]&=\delta_{m,-m'}[(1,0)]+[(\omega,(m+m')e_1)]+[(\omega^3,(m+m')e_1)],  \nonumber \\
[(\omega^3,me_1)]~[(\omega^3,m'e_1)]&=[(\omega,(m+m')e_1)]+[(\omega^3,(m+m')e_1)],  \nonumber \\
[(\omega^k,me_1)]~[(\theta^\ell,0)]&=\delta_{m,0}[(\theta^\ell,0)],  \nonumber \\
[(\theta,0)]~[(\theta,0)]&=[(\theta^2,0)],  \nonumber \\
[(\theta^2,0)]~[(\theta^2,0)]&=[(\theta,0)],  \nonumber \\
[(\theta,0)]~[(\theta^2,0)]&=[(1,0)]+[(\omega,0)]+[(\omega^3,0)]  .
\end{align}
These multiplication rules are direct products of multiplication rules of $T_7$ conjugacy classes and $\mathbb{Z}_7$.

The combinations of allowed 3-point couplings among $[(1,0)]$, $[(\omega^k,me_1)]$ and $[(\theta^\ell,0)]$ are obtained as 
\begin{align}
 &   [(1,0)] ~[(1,0)] ~[(1,0)], \nonumber \\
 &   [(1,0)]~[(\omega,me_1)]~[(\omega^3,-me_1)], \nonumber \\
  &   [(1,0)] ~[(\theta,0)] ~[(\theta^2,0)], \nonumber \\
   &   [(\omega^k,me_1)]~[(\omega^i,m'e_1)]~[(\omega^j,(-m-m')e_1)], \nonumber \\
  &   [(\theta^\ell,0)] ~[(\theta^\ell,0)] ~[(\theta^\ell,0)],
\end{align}
with $k,i,j=1,3$ and $\ell=1,2$.
Similarly, we can obtain higher order couplings.

For example, suppose that three generations of left-handed and right-handed fermions correspond to $[(1,0)]$, $[(\omega,0)]$ and $[(\omega^3,0)]$ and the Higgs mode corresponds to $[(\omega,0)]$.
Then, we can realize the following Yukawa matrix: 
\begin{align}
    Y=
    \begin{pmatrix}
        0 & 0 & * \\
        0 & * & * \\
        * & * & *
    \end{pmatrix}.
\end{align}
This texture can not be derived from group theory.
This texture is very interesting.
In particular, if only the (3,3) entry has a CP phase and the others are real in the quark sector, that can address a axion-less solution for the strong CP problem~\cite{Liang:2025dkm,Kobayashi:2025thd}\footnote{See for the other approaches based on, e.g., a discrete group-like symmetry~\cite{Antusch:2013rla,Feruglio:2023uof,Petcov:2024vph,Penedo:2024gtb} and a six-dimensional theory on $T^2/\mathbb{Z}_3$~\cite{Liang:2024wbb}.}. This texture is also interesting in the lepton sector \cite{Frampton:2002yf,Fritzsch:2011qv}.
The $T^6/T_7$ orbifold models would lead to other interesting textures.
We would study them elsewhere.

\section{Conclusions}
\label{sec:con}

We have studied coupling selection rules due to boundary conditions 
on Abelian and non-Abelian orbifolds in heterotic string theory.
Boundary conditions of closed strings are classified by conjugacy classes of space group elements $(u,v)$, where $u$ are point group elements and $v$ are winding vectors.
The coupling selection rules due to the point group is Abelian on Abelian orbifolds, i.e., invertible.
However, the coupling selection rules due to the point group on non-Abelian orbifolds are determined by the multiplication rules of conjugacy classes of non-Abelian discrete symmetries.
These are non-invertible.
Note that a single element corresponds to a conjugacy class in an Abelian discrete group.
The sublattice $(1-u)\Lambda$ is important in coupling selection rules of winding vectors.
When we have different sublattices $(1-u)\Lambda$ in the twisted sectors, the coupling selection rules among twisted sectors are non-invertible even on Abelian orbifolds as shown by the $T^2/\mathbb{Z}_4$ orbifold.
In general, there are two or more different sublattice structures $(1-u)\Lambda$ in the twisted sectors on non-Abelian orbifolds.

One interesting example of Yukawa matrices was shown on the $T^6/T_7$ orbifold.
The $T^6/T_7$ orbifold models would lead to other interesting textures.
We would study them systematically elsewhere.
Furthermore, we would extend our analysis to other non-Abelian orbifolds.
We leave it for future studies.

\acknowledgments

This work was supported by JSPS KAKENHI Grant Numbers JP23K03375 (T.K.) and JP25H01539 (H.O.).

\bibliography{references}{}

\providecommand{\href}[2]{#2}\begingroup\raggedright\begin{thebibliography}{10}

\bibitem{Kobayashi:2024yqq}
T.~Kobayashi and H.~Otsuka, \emph{{Non-invertible flavor symmetries in magnetized extra dimensions}}, \href{https://doi.org/10.1007/JHEP11(2024)120}{\emph{JHEP} {\bfseries 11} (2024) 120} [\href{https://arxiv.org/abs/2408.13984}{{\ttfamily 2408.13984}}].

\bibitem{Funakoshi:2024uvy}
S.~Funakoshi, T.~Kobayashi and H.~Otsuka, \emph{{Quantum aspects of non-invertible flavor symmetries in intersecting/magnetized D-brane models}}, \href{https://doi.org/10.1007/JHEP04(2025)183}{\emph{JHEP} {\bfseries 04} (2025) 183} [\href{https://arxiv.org/abs/2412.12524}{{\ttfamily 2412.12524}}].

\bibitem{Kobayashi:2024cvp}
T.~Kobayashi, H.~Otsuka and M.~Tanimoto, \emph{{Yukawa textures from non-invertible symmetries}}, \href{https://doi.org/10.1007/JHEP12(2024)117}{\emph{JHEP} {\bfseries 12} (2024) 117} [\href{https://arxiv.org/abs/2409.05270}{{\ttfamily 2409.05270}}].

\bibitem{Kobayashi:2025znw}
T.~Kobayashi, Y.~Nishioka, H.~Otsuka and M.~Tanimoto, \emph{{More about quark Yukawa textures from selection rules without group actions}}, \href{https://doi.org/10.1007/JHEP05(2025)177}{\emph{JHEP} {\bfseries 05} (2025) 177} [\href{https://arxiv.org/abs/2503.09966}{{\ttfamily 2503.09966}}].

\bibitem{Kobayashi:2025ldi}
T.~Kobayashi, H.~Otsuka, M.~Tanimoto and H.~Uchida, \emph{{Lepton mass textures from non-invertible multiplication rules}}, \href{https://doi.org/10.1007/JHEP08(2025)189}{\emph{JHEP} {\bfseries 08} (2025) 189} [\href{https://arxiv.org/abs/2505.07262}{{\ttfamily 2505.07262}}].

\bibitem{Kobayashi:2025cwx}
T.~Kobayashi, H.~Okada and H.~Otsuka, \emph{{Radiative neutrino mass models from non-invertible selection rules}},  \href{https://arxiv.org/abs/2505.14878}{{\ttfamily 2505.14878}}.

\bibitem{Nomura:2025sod}
T.~Nomura and H.~Okada, \emph{{Radiative lepton seesaw model in a non-invertible fusion rule and gauged $B-L$ symmetry}},  \href{https://arxiv.org/abs/2506.16706}{{\ttfamily 2506.16706}}.

\bibitem{Chen:2025awz}
J.~Chen, C.-Q.~Geng, H.~Okada and J.-J.~Wu, \emph{{A radiative lepton model in a non-invertible fusion rule}},  \href{https://arxiv.org/abs/2507.11951}{{\ttfamily 2507.11951}}.

\bibitem{Okada:2025kfm}
H.~Okada and Y.~Shigekami, \emph{{Three-loop induced neutrino mass model in a non-invertible symmetry}},  \href{https://arxiv.org/abs/2507.16198}{{\ttfamily 2507.16198}}.

\bibitem{Jangid:2025krp}
S.~Jangid and H.~Okada, \emph{{A natural realization of inverse seesaw model in a non-invertible selection rule}},  \href{https://arxiv.org/abs/2508.16174}{{\ttfamily 2508.16174}}.

\bibitem{Liang:2025dkm}
Q.~Liang and T.T.~Yanagida, \emph{{Non-invertible symmetry as an axion-less solution to the strong CP problem}}, \href{https://doi.org/10.1016/j.physletb.2025.139706}{\emph{Phys. Lett. B} {\bfseries 868} (2025) 139706} [\href{https://arxiv.org/abs/2505.05142}{{\ttfamily 2505.05142}}].

\bibitem{Kobayashi:2025thd}
T.~Kobayashi, H.~Otsuka and T.T.~Yanagida, \emph{{Non-invertible Symmetry as a Solution to the Strong CP Problem in a GUT-inspired Standard Model}},  \href{https://arxiv.org/abs/2508.12287}{{\ttfamily 2508.12287}}.

\bibitem{Suzuki:2025oov}
M.~Suzuki and L.-X.~Xu, \emph{{Phenomenological implications of a class of non-invertible selection rules}},  \href{https://arxiv.org/abs/2503.19964}{{\ttfamily 2503.19964}}.

\bibitem{Kobayashi:2025lar}
T.~Kobayashi, H.~Mita, H.~Otsuka and R.~Sakuma, \emph{{Matter symmetries in supersymmetric standard models from non-invertible selection rules}},  \href{https://arxiv.org/abs/2506.10241}{{\ttfamily 2506.10241}}.

\bibitem{Suzuki:2025bxg}
M.~Suzuki, L.-X.~Xu and H.Y.~Zhang, \emph{{Spurion Analysis for Non-Invertible Selection Rules from Near-Group Fusions}},  \href{https://arxiv.org/abs/2508.14970}{{\ttfamily 2508.14970}}.

\bibitem{Gomes:2023ahz}
P.R.S.~Gomes, \emph{{An introduction to higher-form symmetries}}, \href{https://doi.org/10.21468/SciPostPhysLectNotes.74}{\emph{SciPost Phys. Lect. Notes} {\bfseries 74} (2023) 1} [\href{https://arxiv.org/abs/2303.01817}{{\ttfamily 2303.01817}}].

\bibitem{Schafer-Nameki:2023jdn}
S.~Schafer-Nameki, \emph{{ICTP lectures on (non-)invertible generalized symmetries}}, \href{https://doi.org/10.1016/j.physrep.2024.01.007}{\emph{Phys. Rept.} {\bfseries 1063} (2024) 1} [\href{https://arxiv.org/abs/2305.18296}{{\ttfamily 2305.18296}}].

\bibitem{Bhardwaj:2023kri}
L.~Bhardwaj, L.E.~Bottini, L.~Fraser-Taliente, L.~Gladden, D.S.W.~Gould, A.~Platschorre et~al., \emph{{Lectures on generalized symmetries}}, \href{https://doi.org/10.1016/j.physrep.2023.11.002}{\emph{Phys. Rept.} {\bfseries 1051} (2024) 1} [\href{https://arxiv.org/abs/2307.07547}{{\ttfamily 2307.07547}}].

\bibitem{Shao:2023gho}
S.-H.~Shao, \emph{{What's Done Cannot Be Undone: TASI Lectures on Non-Invertible Symmetries}},  \href{https://arxiv.org/abs/2308.00747}{{\ttfamily 2308.00747}}.

\bibitem{Choi:2022jqy}
Y.~Choi, H.T.~Lam and S.-H.~Shao, \emph{{Noninvertible Global Symmetries in the Standard Model}}, \href{https://doi.org/10.1103/PhysRevLett.129.161601}{\emph{Phys. Rev. Lett.} {\bfseries 129} (2022) 161601} [\href{https://arxiv.org/abs/2205.05086}{{\ttfamily 2205.05086}}].

\bibitem{Cordova:2022fhg}
C.~Cordova, S.~Hong, S.~Koren and K.~Ohmori, \emph{{Neutrino Masses from Generalized Symmetry Breaking}}, \href{https://doi.org/10.1103/PhysRevX.14.031033}{\emph{Phys. Rev. X} {\bfseries 14} (2024) 031033} [\href{https://arxiv.org/abs/2211.07639}{{\ttfamily 2211.07639}}].

\bibitem{Cordova:2022ieu}
C.~Cordova and K.~Ohmori, \emph{{Noninvertible Chiral Symmetry and Exponential Hierarchies}}, \href{https://doi.org/10.1103/PhysRevX.13.011034}{\emph{Phys. Rev. X} {\bfseries 13} (2023) 011034} [\href{https://arxiv.org/abs/2205.06243}{{\ttfamily 2205.06243}}].

\bibitem{Cordova:2024ypu}
C.~Cordova, S.~Hong and S.~Koren, \emph{{Non-Invertible Peccei-Quinn Symmetry and the Massless Quark Solution to the Strong CP Problem}},  \href{https://arxiv.org/abs/2402.12453}{{\ttfamily 2402.12453}}.

\bibitem{Delgado:2024pcv}
A.~Delgado and S.~Koren, \emph{{Non-invertible Peccei-Quinn symmetry, natural 2HDM alignment, and the visible axion}}, \href{https://doi.org/10.1007/JHEP02(2025)178}{\emph{JHEP} {\bfseries 02} (2025) 178} [\href{https://arxiv.org/abs/2412.05362}{{\ttfamily 2412.05362}}].

\bibitem{Dixon:1985jw}
L.J.~Dixon, J.A.~Harvey, C.~Vafa and E.~Witten, \emph{{Strings on Orbifolds}}, \href{https://doi.org/10.1016/0550-3213(85)90593-0}{\emph{Nucl. Phys. B} {\bfseries 261} (1985) 678}.

\bibitem{Dixon:1986jc}
L.J.~Dixon, J.A.~Harvey, C.~Vafa and E.~Witten, \emph{{Strings on Orbifolds. 2.}}, \href{https://doi.org/10.1016/0550-3213(86)90287-7}{\emph{Nucl. Phys. B} {\bfseries 274} (1986) 285}.

\bibitem{Ibanez:1986tp}
L.E.~Ibanez, H.P.~Nilles and F.~Quevedo, \emph{{Orbifolds and Wilson Lines}}, \href{https://doi.org/10.1016/0370-2693(87)90066-9}{\emph{Phys. Lett. B} {\bfseries 187} (1987) 25}.

\bibitem{Ibanez:1987sn}
L.E.~Ibanez, J.E.~Kim, H.P.~Nilles and F.~Quevedo, \emph{{Orbifold Compactifications with Three Families of SU(3) x SU(2) x U(1)**n}}, \href{https://doi.org/10.1016/0370-2693(87)90255-3}{\emph{Phys. Lett. B} {\bfseries 191} (1987) 282}.

\bibitem{Kobayashi:2004ud}
T.~Kobayashi, S.~Raby and R.-J.~Zhang, \emph{{Constructing 5-D orbifold grand unified theories from heterotic strings}}, \href{https://doi.org/10.1016/j.physletb.2004.04.058}{\emph{Phys. Lett. B} {\bfseries 593} (2004) 262} [\href{https://arxiv.org/abs/hep-ph/0403065}{{\ttfamily hep-ph/0403065}}].

\bibitem{Kobayashi:2004ya}
T.~Kobayashi, S.~Raby and R.-J.~Zhang, \emph{{Searching for realistic 4d string models with a Pati-Salam symmetry: Orbifold grand unified theories from heterotic string compactification on a Z(6) orbifold}}, \href{https://doi.org/10.1016/j.nuclphysb.2004.10.035}{\emph{Nucl. Phys. B} {\bfseries 704} (2005) 3} [\href{https://arxiv.org/abs/hep-ph/0409098}{{\ttfamily hep-ph/0409098}}].

\bibitem{Buchmuller:2005jr}
W.~Buchmuller, K.~Hamaguchi, O.~Lebedev and M.~Ratz, \emph{{Supersymmetric standard model from the heterotic string}}, \href{https://doi.org/10.1103/PhysRevLett.96.121602}{\emph{Phys. Rev. Lett.} {\bfseries 96} (2006) 121602} [\href{https://arxiv.org/abs/hep-ph/0511035}{{\ttfamily hep-ph/0511035}}].

\bibitem{Buchmuller:2006ik}
W.~Buchmuller, K.~Hamaguchi, O.~Lebedev and M.~Ratz, \emph{{Supersymmetric Standard Model from the Heterotic String (II)}}, \href{https://doi.org/10.1016/j.nuclphysb.2007.06.028}{\emph{Nucl. Phys. B} {\bfseries 785} (2007) 149} [\href{https://arxiv.org/abs/hep-th/0606187}{{\ttfamily hep-th/0606187}}].

\bibitem{Lebedev:2006kn}
O.~Lebedev, H.P.~Nilles, S.~Raby, S.~Ramos-Sanchez, M.~Ratz, P.K.S.~Vaudrevange et~al., \emph{{A Mini-landscape of exact MSSM spectra in heterotic orbifolds}}, \href{https://doi.org/10.1016/j.physletb.2006.12.012}{\emph{Phys. Lett. B} {\bfseries 645} (2007) 88} [\href{https://arxiv.org/abs/hep-th/0611095}{{\ttfamily hep-th/0611095}}].

\bibitem{Lebedev:2007hv}
O.~Lebedev, H.P.~Nilles, S.~Raby, S.~Ramos-Sanchez, M.~Ratz, P.K.S.~Vaudrevange et~al., \emph{{The Heterotic Road to the MSSM with R parity}}, \href{https://doi.org/10.1103/PhysRevD.77.046013}{\emph{Phys. Rev. D} {\bfseries 77} (2008) 046013} [\href{https://arxiv.org/abs/0708.2691}{{\ttfamily 0708.2691}}].

\bibitem{Hamidi:1986vh}
S.~Hamidi and C.~Vafa, \emph{{Interactions on Orbifolds}}, \href{https://doi.org/10.1016/0550-3213(87)90006-X}{\emph{Nucl. Phys. B} {\bfseries 279} (1987) 465}.

\bibitem{Dixon:1986qv}
L.J.~Dixon, D.~Friedan, E.J.~Martinec and S.H.~Shenker, \emph{{The Conformal Field Theory of Orbifolds}}, \href{https://doi.org/10.1016/0550-3213(87)90676-6}{\emph{Nucl. Phys. B} {\bfseries 282} (1987) 13}.

\bibitem{Burwick:1990tu}
T.T.~Burwick, R.K.~Kaiser and H.F.~Muller, \emph{{General Yukawa couplings of strings on Z(N) orbifolds}}, \href{https://doi.org/10.1016/0550-3213(91)90491-F}{\emph{Nucl. Phys. B} {\bfseries 355} (1991) 689}.

\bibitem{Choi:2007nb}
K.-S.~Choi and T.~Kobayashi, \emph{{Higher order couplings from heterotic orbifold theory}}, \href{https://doi.org/10.1016/j.nuclphysb.2008.01.016}{\emph{Nucl. Phys. B} {\bfseries 797} (2008) 295} [\href{https://arxiv.org/abs/0711.4894}{{\ttfamily 0711.4894}}].

\bibitem{Font:1988nc}
A.~Font, L.E.~Ibanez, H.P.~Nilles and F.~Quevedo, \emph{{On the Concept of Naturalness in String Theories}}, \href{https://doi.org/10.1016/0370-2693(88)91760-1}{\emph{Phys. Lett. B} {\bfseries 213} (1988) 274}.

\bibitem{CaboBizet:2013gns}
N.G.~Cabo~Bizet, T.~Kobayashi, D.K.~Mayorga~Pena, S.L.~Parameswaran, M.~Schmitz and I.~Zavala, \emph{{R-charge Conservation and More in Factorizable and Non-Factorizable Orbifolds}}, \href{https://doi.org/10.1007/JHEP05(2013)076}{\emph{JHEP} {\bfseries 05} (2013) 076} [\href{https://arxiv.org/abs/1301.2322}{{\ttfamily 1301.2322}}].

\bibitem{Nilles:2013lda}
H.P.~Nilles, S.~Ramos-S{\'a}nchez, M.~Ratz and P.K.S.~Vaudrevange, \emph{{A note on discrete $R$ symmetries in $\mathbb{Z}_{6}$-II orbifolds with Wilson lines}}, \href{https://doi.org/10.1016/j.physletb.2013.09.041}{\emph{Phys. Lett. B} {\bfseries 726} (2013) 876} [\href{https://arxiv.org/abs/1308.3435}{{\ttfamily 1308.3435}}].

\bibitem{CaboBizet:2013hms}
N.G.~Cabo~Bizet, T.~Kobayashi, D.K.~Mayorga~Pena, S.L.~Parameswaran, M.~Schmitz and I.~Zavala, \emph{{Discrete R-symmetries and Anomaly Universality in Heterotic Orbifolds}}, \href{https://doi.org/10.1007/JHEP02(2014)098}{\emph{JHEP} {\bfseries 02} (2014) 098} [\href{https://arxiv.org/abs/1308.5669}{{\ttfamily 1308.5669}}].

\bibitem{Dijkgraaf:1987vp}
R.~Dijkgraaf, E.P.~Verlinde and H.L.~Verlinde, \emph{{C = 1 Conformal Field Theories on Riemann Surfaces}}, \href{https://doi.org/10.1007/BF01224132}{\emph{Commun. Math. Phys.} {\bfseries 115} (1988) 649}.

\bibitem{Kobayashi:2006wq}
T.~Kobayashi, H.P.~Nilles, F.~Ploger, S.~Raby and M.~Ratz, \emph{{Stringy origin of non-Abelian discrete flavor symmetries}}, \href{https://doi.org/10.1016/j.nuclphysb.2007.01.018}{\emph{Nucl. Phys. B} {\bfseries 768} (2007) 135} [\href{https://arxiv.org/abs/hep-ph/0611020}{{\ttfamily hep-ph/0611020}}].

\bibitem{Beye:2014nxa}
F.~Beye, T.~Kobayashi and S.~Kuwakino, \emph{{Gauge Origin of Discrete Flavor Symmetries in Heterotic Orbifolds}}, \href{https://doi.org/10.1016/j.physletb.2014.07.058}{\emph{Phys. Lett. B} {\bfseries 736} (2014) 433} [\href{https://arxiv.org/abs/1406.4660}{{\ttfamily 1406.4660}}].

\bibitem{Thorngren:2021yso}
R.~Thorngren and Y.~Wang, \emph{{Fusion category symmetry. Part II. Categoriosities at c = 1 and beyond}}, \href{https://doi.org/10.1007/JHEP07(2024)051}{\emph{JHEP} {\bfseries 07} (2024) 051} [\href{https://arxiv.org/abs/2106.12577}{{\ttfamily 2106.12577}}].

\bibitem{Heckman:2024obe}
J.J.~Heckman, J.~McNamara, M.~Montero, A.~Sharon, C.~Vafa and I.~Valenzuela, \emph{{On the Fate of Stringy Non-Invertible Symmetries}},  \href{https://arxiv.org/abs/2402.00118}{{\ttfamily 2402.00118}}.

\bibitem{Kaidi:2024wio}
J.~Kaidi, Y.~Tachikawa and H.Y.~Zhang, \emph{{On a class of selection rules without group actions in field theory and string theory}}, \href{https://doi.org/10.21468/SciPostPhys.17.6.169}{\emph{SciPost Phys.} {\bfseries 17} (2024) 169} [\href{https://arxiv.org/abs/2402.00105}{{\ttfamily 2402.00105}}].

\bibitem{Kobayashi:1990mc}
T.~Kobayashi and N.~Ohtsubo, \emph{{Yukawa Coupling Condition of $Z(N$) Orbifold Models}}, \href{https://doi.org/10.1016/0370-2693(90)90671-R}{\emph{Phys. Lett. B} {\bfseries 245} (1990) 441}.

\bibitem{Kobayashi:1991rp}
T.~Kobayashi and N.~Ohtsubo, \emph{{Geometrical aspects of Z(N) orbifold phenomenology}}, \href{https://doi.org/10.1142/S0217751X94000054}{\emph{Int. J. Mod. Phys. A} {\bfseries 9} (1994) 87}.

\bibitem{Kobayashi:1995py}
T.~Kobayashi, \emph{{Selection rules for nonrenormalizable couplings in superstring theories}}, \href{https://doi.org/10.1016/0370-2693(95)00643-Y}{\emph{Phys. Lett. B} {\bfseries 354} (1995) 264} [\href{https://arxiv.org/abs/hep-ph/9504371}{{\ttfamily hep-ph/9504371}}].

\bibitem{Cvetic:1987qx}
M.~Cvetic, \emph{{Suppression of Nonrenormalizable Terms in the Effective Superpotential for (Blownup) Orbifold Compactification}}, \href{https://doi.org/10.1103/PhysRevLett.59.1795}{\emph{Phys. Rev. Lett.} {\bfseries 59} (1987) 1795}.

\bibitem{Kobayashi:2011cw}
T.~Kobayashi, S.L.~Parameswaran, S.~Ramos-Sanchez and I.~Zavala, \emph{{Revisiting Coupling Selection Rules in Heterotic Orbifold Models}}, \href{https://doi.org/10.1007/JHEP12(2012)049}{\emph{JHEP} {\bfseries 05} (2012) 008} [\href{https://arxiv.org/abs/1107.2137}{{\ttfamily 1107.2137}}].

\bibitem{Dong:2025pah}
J.~Dong, T.~Kobayashi, R.~Nishida, S.~Nishimura and H.~Otsuka, \emph{{Coupling Selection Rules in Heterotic Calabi-Yau Compactifications}},  \href{https://arxiv.org/abs/2504.09773}{{\ttfamily 2504.09773}}.

\bibitem{Inoue:1987ak}
K.~Inoue, M.~Sakamoto and H.~Takano, \emph{{NONABELIAN ORBIFOLDS}}, \href{https://doi.org/10.1143/PTP.78.908}{\emph{Prog. Theor. Phys.} {\bfseries 78} (1987) 908}.

\bibitem{Inoue:1988ki}
K.~Inoue, S.~Nima and H.~Takano, \emph{{Zero Mode and Modular Invariance in String on Nonabelian Orbifold}}, \href{https://doi.org/10.1143/PTP.80.881}{\emph{Prog. Theor. Phys.} {\bfseries 80} (1988) 881}.

\bibitem{Inoue:1990ci}
K.~Inoue and S.~Nima, \emph{{String interactions on nonAbelian orbifold}}, \href{https://doi.org/10.1143/PTP.84.702}{\emph{Prog. Theor. Phys.} {\bfseries 84} (1990) 702}.

\bibitem{Konopka:2012gy}
S.J.H.~Konopka, \emph{{Non Abelian orbifold compactifications of the heterotic string}}, \href{https://doi.org/10.1007/JHEP07(2013)023}{\emph{JHEP} {\bfseries 07} (2013) 023} [\href{https://arxiv.org/abs/1210.5040}{{\ttfamily 1210.5040}}].

\bibitem{Fischer:2012qj}
M.~Fischer, M.~Ratz, J.~Torrado and P.K.S.~Vaudrevange, \emph{{Classification of symmetric toroidal orbifolds}}, \href{https://doi.org/10.1007/JHEP01(2013)084}{\emph{JHEP} {\bfseries 01} (2013) 084} [\href{https://arxiv.org/abs/1209.3906}{{\ttfamily 1209.3906}}].

\bibitem{Fischer:2013qza}
M.~Fischer, S.~Ramos-Sanchez and P.K.S.~Vaudrevange, \emph{{Heterotic non-Abelian orbifolds}}, \href{https://doi.org/10.1007/JHEP07(2013)080}{\emph{JHEP} {\bfseries 07} (2013) 080} [\href{https://arxiv.org/abs/1304.7742}{{\ttfamily 1304.7742}}].

\bibitem{Funakoshi:2025lxs}
S.~Funakoshi, Y.~Koga and H.~Otsuka, \emph{{Classification of Modular Symmetries in Non-Supersymmetric Heterotic String theories}},  \href{https://arxiv.org/abs/2503.23741}{{\ttfamily 2503.23741}}.

\bibitem{Hernandez-Segura:2025sfr}
M.~Hernandez-Segura and S.~Ramos-Sanchez, \emph{{Non-Abelian orbifolds of the SO(32) heterotic string}},  \href{https://arxiv.org/abs/2506.08370}{{\ttfamily 2506.08370}}.

\bibitem{Ishimori:2010au}
H.~Ishimori, T.~Kobayashi, H.~Ohki, Y.~Shimizu, H.~Okada and M.~Tanimoto, \emph{{Non-Abelian Discrete Symmetries in Particle Physics}}, \href{https://doi.org/10.1143/PTPS.183.1}{\emph{Prog. Theor. Phys. Suppl.} {\bfseries 183} (2010) 1} [\href{https://arxiv.org/abs/1003.3552}{{\ttfamily 1003.3552}}].

\bibitem{Kobayashi:2022moq}
T.~Kobayashi, H.~Ohki, H.~Okada, Y.~Shimizu and M.~Tanimoto, \emph{{An Introduction to Non-Abelian Discrete Symmetries for Particle Physicists}} (1, 2022), \href{https://doi.org/10.1007/978-3-662-64679-3}{10.1007/978-3-662-64679-3}.

\bibitem{Ramos-Sanchez:2018edc}
S.~Ramos-S{\'a}nchez and P.K.S.~Vaudrevange, \emph{{Note on the space group selection rule for closed strings on orbifolds}}, \href{https://doi.org/10.1007/JHEP01(2019)055}{\emph{JHEP} {\bfseries 01} (2019) 055} [\href{https://arxiv.org/abs/1811.00580}{{\ttfamily 1811.00580}}].

\bibitem{Dong:2025jra}
J.~Dong, T.~Jeric, T.~Kobayashi, R.~Nishida and H.~Otsuka, \emph{{On discrete gauging and non-invertible selection rules}},  \href{https://arxiv.org/abs/2507.02375}{{\ttfamily 2507.02375}}.

\bibitem{Antusch:2013rla}
S.~Antusch, M.~Holthausen, M.A.~Schmidt and M.~Spinrath, \emph{{Solving the Strong CP Problem with Discrete Symmetries and the Right Unitarity Triangle}}, \href{https://doi.org/10.1016/j.nuclphysb.2013.10.028}{\emph{Nucl. Phys. B} {\bfseries 877} (2013) 752} [\href{https://arxiv.org/abs/1307.0710}{{\ttfamily 1307.0710}}].

\bibitem{Feruglio:2023uof}
F.~Feruglio, A.~Strumia and A.~Titov, \emph{{Modular invariance and the QCD angle}}, \href{https://doi.org/10.1007/JHEP07(2023)027}{\emph{JHEP} {\bfseries 07} (2023) 027} [\href{https://arxiv.org/abs/2305.08908}{{\ttfamily 2305.08908}}].

\bibitem{Petcov:2024vph}
S.T.~Petcov and M.~Tanimoto, \emph{{$A_4$ modular invariance and the strong CP problem}}, \href{https://doi.org/10.1140/epjc/s10052-024-13272-w}{\emph{Eur. Phys. J. C} {\bfseries 84} (2024) 914} [\href{https://arxiv.org/abs/2404.00858}{{\ttfamily 2404.00858}}].

\bibitem{Penedo:2024gtb}
J.T.~Penedo and S.T.~Petcov, \emph{{Finite modular symmetries and the strong CP problem}}, \href{https://doi.org/10.1007/JHEP10(2024)172}{\emph{JHEP} {\bfseries 10} (2024) 172} [\href{https://arxiv.org/abs/2404.08032}{{\ttfamily 2404.08032}}].

\bibitem{Liang:2024wbb}
Q.~Liang, R.~Okabe and T.T.~Yanagida, \emph{{Three-zero texture of quark-mass matrices as a solution to the strong CP problem}}, \href{https://doi.org/10.1016/j.physletb.2024.139123}{\emph{Phys. Lett. B} {\bfseries 859} (2024) 139123} [\href{https://arxiv.org/abs/2408.12146}{{\ttfamily 2408.12146}}].

\bibitem{Frampton:2002yf}
P.H.~Frampton, S.L.~Glashow and D.~Marfatia, \emph{{Zeroes of the neutrino mass matrix}}, \href{https://doi.org/10.1016/S0370-2693(02)01817-8}{\emph{Phys. Lett. B} {\bfseries 536} (2002) 79} [\href{https://arxiv.org/abs/hep-ph/0201008}{{\ttfamily hep-ph/0201008}}].

\bibitem{Fritzsch:2011qv}
H.~Fritzsch, Z.-z.~Xing and S.~Zhou, \emph{{Two-zero Textures of the Majorana Neutrino Mass Matrix and Current Experimental Tests}}, \href{https://doi.org/10.1007/JHEP09(2011)083}{\emph{JHEP} {\bfseries 09} (2011) 083} [\href{https://arxiv.org/abs/1108.4534}{{\ttfamily 1108.4534}}].

\end{thebibliography}\endgroup
\bibliographystyle{JHEP}

\end{document}